\documentclass[12pt,preprint]{aastex}
\usepackage{mathptmx}
\usepackage{graphicx}

\newcommand{\nuclei}[2]{\ensuremath{\mathrm{^{#1}#2}}}
\newcommand{\carbon}[1][12]{\nuclei{#1}{C}}
\newcommand{\nitrogen}[1][14]{\nuclei{#1}{N}}
\newcommand{\oxygen}[1][16]{\nuclei{#1}{O}}
\newcommand{\neon}[1][20]{\nuclei{#1}{Ne}}
\newcommand{\iron}[1][56]{\nuclei{#1}{Fe}}
\newcommand{\cobalt}[1][56]{\nuclei{#1}{Co}}
\newcommand{\nickel}[1][58]{\nuclei{#1}{Ni}}

\newcommand{\OH}{\ensuremath{\left[\mathrm{O/H}\right]}}

\newcommand{\YO}{\ensuremath{Y_{\mathrm{O}}}}
\newcommand{\YFe}{\ensuremath{Y_{\mathrm{Fe}}}}
\newcommand{\YC}{\ensuremath{Y_{\mathrm{C}}}}
\newcommand{\YN}{\ensuremath{Y_{\mathrm{N}}}}
\newcommand{\YNe}{\ensuremath{Y_{\mathrm{Ne}}}}

\newcommand{\Ye}{\ensuremath{Y_{e}}}
\newcommand{\OFe}{\ensuremath{\left[\mathrm{O/Fe}\right]}}
\newcommand{\FeH}{\ensuremath{\left[\mathrm{Fe/H}\right]}}
\newcommand{\Msun}{${\rm M}_\odot$}     
\newcommand{\drp}{\ensuremath{\Delta m_{15}}}     
\newcommand{\Sdot}{\ensuremath{\dot{S}}}     
\newcommand{\ugriz}{$u_Mg_Mr_Mi_Mz_M$}
\newcommand{\griz}{$g_Mr_Mi_Mz_M$}
\begin{document}
\title{The effect of progenitor age and metallicity on luminosity and $^{56}$Ni yield in Type Ia supernovae}

\author{D.~A.~Howell\altaffilmark{1,2,3},
M.~Sullivan\altaffilmark{4,1},
E.~F.~Brown\altaffilmark{5},
A.~Conley\altaffilmark{1},
D.~Le~Borgne\altaffilmark{6},
E.~Y.~Hsiao\altaffilmark{7},
P.~Astier\altaffilmark{8}, 
D.~Balam\altaffilmark{7}, 
C.~Balland\altaffilmark{8,9},
S.~Basa\altaffilmark{10},
R.~G.~Carlberg\altaffilmark{1},
D.~Fouchez\altaffilmark{11}, 
J.~Guy\altaffilmark{8}, 
D.~Hardin\altaffilmark{8}, 
I.~M.~Hook\altaffilmark{4}, 
R.~Pain\altaffilmark{8}, 
K.~Perrett\altaffilmark{1}, 
C.~J.~Pritchet\altaffilmark{7},
N.~Regnault\altaffilmark{8}, 
S.~Baumont\altaffilmark{8},
J. Le Du\altaffilmark{11},
C. Lidman\altaffilmark{12},
S.~Perlmutter\altaffilmark{13,14},
N. Suzuki\altaffilmark{13},
E.~S.~Walker\altaffilmark{4},
J.~C.~Wheeler\altaffilmark{15}
}
                                                          
\altaffiltext{1}{Department of Astronomy and Astrophysics, University of
Toronto, 50 St. George Street, Toronto, ON M5S 3H8, Canada}
\altaffiltext{2}{Las Cumbres Observatory Global Telescope Network,
  6740 Cortona Dr., Suite 102, Goleta, CA 93117}
\altaffiltext{3}{Department of Physics, University of California,
  Santa Barbara, Broida Hall, Mail Code 9530, Santa Barbara, CA 93106-9530}
\altaffiltext{4}{University of Oxford Astrophysics, Denys Wilkinson
  Building, Keble Road, Oxford OX1 3RH, UK}
\altaffiltext{5}{Department of Physics \& Astronomy, National
Superconducting Cyclotron Laboratory, and the Joint Institute for
Nuclear Astrophysics, Michigan State University, East  Lansing, MI 48824}
\altaffiltext{6}{Institut d'Astrophysique de Paris, UMR7095 CNRS, UPMC, 98 bis Boulevard Arago, 75014 Paris, France}
\altaffiltext{7}{Department of Physics and Astronomy, University of
Victoria, PO Box 3055, Victoria, BC V8W 3P6, Canada}
\altaffiltext{8}{LPNHE, CNRS-IN2P3 and University of
Paris VI \& VII, 75005 Paris, France}
\altaffiltext{9}{Univ. Paris-Sud, Orsay, F-91405, France}
\altaffiltext{10}{Laboratoire d'Astrophysique de Marseille, P\^ole de
  l'Etoile Site de Ch\^ateau-Gombert, 38, rue Fr\'ed\'eric Joliot-Curie, 13388 Marseille cedex 13, France}
\altaffiltext{11}{CPPM, CNRS-Luminy, Case 907, 13288 Marseille Cedex 9, France}
\altaffiltext{12}{ESO, Alonso de Cordova, 3107, Vitacura Casilla 19001,
  Santiago 19, Chile}
\altaffiltext{13}{Lawrence Berkeley National Laboratory, Mail Stop
  50-232, Lawrence Berkeley National Laboratory, 1 Cyclotron Road, Berkeley CA 94720 USA}
\altaffiltext{14}{Department of Physics, University of California,
366 LeConte Hall MC 7300,
Berkeley, CA 94720-7300 USA}
\altaffiltext{15}{Department of Astronomy, University of Texas, RLM 5.208,
Austin, TX 78712-1081 USA}



\begin{abstract}
  \citeauthor{2003ApJ...590L..83T} found that metallicity variations
  could theoretically account for a 25\% variation in the mass of
  \nickel[56] synthesized in Type Ia supernovae (SNe Ia), and thus
  account for a large fraction of the scatter in observed SN Ia
  luminosities.  Higher-metallicity progenitors are more neutron-rich,
  producing more stable burning products relative to radioactive
  \nickel[56].  We develop a new method for estimating bolometric
  luminosity and \nickel[56] yield in SNe Ia and use it to test
  the theory with data from the Supernova Legacy Survey.  We find that
  the average \nickel[56] yield does drop in SNe Ia from high
  metallicity environments, but the theory can only account for
  7\%--10\% of the dispersion in SN Ia \nickel[56] mass, and thus
  luminosity.  This is because the effect is dominant at metallicities
  significantly above solar, whereas we find that SN hosts have
  predominantly subsolar or only moderately above-solar metallicities.
  We also show that allowing for changes in O/Fe with the metallicity
  [Fe/H] does not have a major effect on the theoretical prediction of
  \citeauthor{2003ApJ...590L..83T}, so long as one is using the O/H as
  the independent variable.  Age may have a greater effect than
  metallicity -- we find that the luminosity weighted age of the host
  galaxy is correlated with \nickel[56] yield, and thus more massive
  progenitors give rise to more luminous explosions.  This is hard to
  understand if most SNe Ia explode when the primaries reach the
  Chandrasekhar mass.  Finally, we test the findings of
  \citeauthor{2008arXiv0805.4360G}, that the residuals of SNe Ia from
  the Hubble diagram are correlated with host galaxy metallicity, and
  we find no such correlation.

\end{abstract}


\section{Introduction}
Type Ia supernovae (SNe Ia) make excellent cosmological standard
candles \citep{1998AJ....116.1009R,1999ApJ...517..565P}, though only
after they are empirically calibrated based on their lightcurve shape
\citep{1993ApJ...413L.105P} and color \citep{1996ApJ...473...88R}.
There is consensus that the luminosity of Type Ia supernovae (SNe Ia)
arises from the radioactive decay of \nickel[56] to \cobalt[56], and
finally to \iron[56] \citep{1967CaJPh..45.2315T,1969ApJ...157..623C}.
Therefore, the primary variable controlling SN luminosity is the
amount of \nickel[56] synthesized during the explosion, though as
second-order effects the location of the \nickel[56]
\citep[e.g.][]{1995PhR...256...53B, 1997ApJ...485..812N,
  2000ApJ...530..744P}, the opacity of the overlying material
\citep[e.g.][]{1993A&A...270..223K, 2000ApJ...530..757P,
  2001ApJ...547..988M,2007ApJ...656..661K}, and asymmetries in the
ejecta \citep{2001ApJ...556..302H,2004ApJ...610..876K} can
redistribute the energy in wavelength, time, or space.  However,
because SNe Ia are largely spherical
\citep{1996ApJ...467..435W,2005ApJ...632..450L}, the bulk of
\nickel[56] is located near the center, and simple assumptions about
the opacity are a reasonable approximation, Arnett's Rule
\citep{1979ApJ...230L..37A,1982ApJ...253..785A} can be used to
empirically estimate the \nickel[56] mass from the optical-infrared
lightcurve to an accuracy of 5 -- 15\%
\citep{2006A&A...460..793S,2006A&A...453..229B}.

The variables that control the yield of \nickel[56] in a SN Ia are not
well understood, although several factors are known to be important.
Iron-peak elements are produced when burning occurs in a state of
nuclear statistical equilibrium, which happens at the highest
densities near the center of the white dwarf.  If this burning occurs
in a neutron-rich region, more neutron-rich, stable, iron-peak elements
like \nickel[58] and \iron[54] will be
produced relative to \nickel[56].  Since main-sequence stars with higher
metallicity abundances produce white dwarfs with more of the
neutron-rich nuclides \neon[22] and \iron[56], \citet[][hereafter
TBT03]{2003ApJ...590L..83T} hypothesized that high metallicity
progenitors will produce less \nickel[56] and therefore less luminous
SNe Ia.

In the absence of any identified SN Ia progenitor, the best one could
hope for is to see traces of the progenitor's metallicity in the
unburned outer layers of the SN by looking at Type Ia spectra from
only a few days after the explosion in the UV
\citep{1998ApJ...495..617H,2001ApJ...557..266L}.  Unfortunately the
difficulty of obtaining UV spectra has prevented the gathering of a
statistically significant sample of spectra from the earliest phases.
In the case of local SNe~Ia it has not been possible to schedule HST
quickly enough \citep{2008arXiv0803.1181F,2008arXiv0803.0871S}.  Another option
is to observe higher redshift SNe Ia in the restframe UV, though
their faintness at such early epochs relative to their host galaxy
makes observations earlier than one week before maximum light difficult
\citep{2008ApJ...674...51E}.  More importantly, even when UV spectra have been
obtained their theoretical interpretation is not
straightforward \citep{2008ApJ...674...51E}.

Since a true progenitor metallicity has so far been unobtainable, the
next best thing is to determine the average metallicity in the
environment of the SN.  The two most commonly used approaches have
been to measure spectroscopic line indices
\citep{2000AJ....120.1479H,2008arXiv0805.4360G} or ratios
\citep{2005ApJ...634..210G} to determine a global average host galaxy
metallicity, or to take advantage of the fact that galaxies often have
metallicity gradients, and study projected galactocentric offset as a
proxy for metallicity \citep{1997ApJ...483L..29W, 2000ApJ...542..588I,
  2005ApJ...634..210G}.  The drawbacks of the spectroscopic line
techniques are that they require complicated corrections for
contamination by emission, and that the interpretation of the line
ratios as metallicity requires mapping observations to a
sometimes-incongruous grid of theoretical models
\citep[e.g.][]{2000AJ....119.1645T, 2002ApJS..142...35K,
  2008arXiv0805.4360G}.  Furthermore, the line ratios only represent
the current average metallicity -- the metallicity when the SN Ia
progenitor formed may have been different.  In addition, high
signal-to-noise ratio spectra are required, and important lines like H$\alpha$ are
redshifted into the IR at $z>0.5$.  These factors limit the possible
sample size of such a data set.  The drawbacks of the offset technique
are that projection effects can be confusing
\citep{1997ApJ...483L..29W,1997AJ....113..197V}, as is differential
reddening between SNe near and far from galaxy centers
\citep{1998ApJ...502..177H}, and Malmquist-like effects
\citep{2000ApJ...530..166H}.  Further complications include that
different galaxy types have different metallicity gradients, and the
current location of the SN may not be representative of where the
progenitor formed.

With the exception of \citet{2008arXiv0805.4360G}, hereafter G08, none
of the above studies found a significant correlation between
metallicity and SN Ia properties, although it is not clear whether
this is caused by the lack of an effect or whether the techniques used
so far do not have the required level of precision.  G08 compared
absorption-line strengths of 29 E/S0 galaxies which hosted SNe Ia to a
grid of models and found that galaxies with high iron abundances host
less luminous supernovae.  They went on to conclude that age played a
greater role, since galaxies with characteristic ages greater than 5
Gyr produce $\sim 1$ mag fainter SNe Ia in $V$.  Their most
controversial finding is that the MLCS lightcurve fitter could not
fully correct for SN luminosities in different metallicity
environments; when Hubble residuals were correlated with metallicity,
they found a significant trend.  But since these findings are based on
a small sample size, rely on an uncertain metallicity determination
techniques, and rely entirely on a lightcurve fitter
\citep[MLCS2k2;][]{2007ApJ...659..122J} that has recently shown
spurious systematic trends with Hubble residuals
\citep{2007ApJ...664L..13C}, they are worth scrutinizing with
independent techniques and a much larger sample size.

Here we use an independent approach not previously applied to testing
SN luminosity-metallicity relations --- we estimate the host galaxy
metallicity from its mass using spectral energy distribution (SED)
fits to the host galaxy photometry and the \citet{2004ApJ...613..898T}
mass-metallicity relation.

Another discriminator of the metallicity effect is that it
may cause luminosity differences as a function of redshift, since
cosmic metallicity is increasing with time
\citep[e.g.][]{2005ApJ...635..260S}.  This is in contrast to many
other possible contributors to SN Ia luminosity (for example, white
dwarf core density) which would not be expected to change with
redshift.  We perform the first tests of the effect of inferred metallicity on
SN Ia nickel mass with high redshift supernovae, using data from the
Supernova Legacy Survey (SNLS).

In this paper we first update the TBT03 SN Ia 
metallicity-\nickel[56] mass relation to account for the fact that 
O/Fe can vary as a function of Fe/H (\S 2).  Then we test the
theory using data from the SNLS.  To do this, in \S 3.1 we estimate host
galaxy masses from PEGASE-2 model fits to \ugriz\
(subscripts denote Megacam filters) host galaxy photometry.  In \S 3.2
we use galaxy mass-metallicity relations to estimate host
metallicities.  In \S 3.3 we determine SN \nickel[56] mass from SNLS
photometry.  Finally, we test the theory with the data in \S 4 and
discuss the implications in \S 5.

\section{The Variation of \nickel[56] Mass with Progenitor Metallicity}
\label{theorymod}

TBT03 note that a white dwarf's \neon[22] abundance should trace its
progenitor star's O abundance since C, N, and O are converted to
\nitrogen\ during hydrogen burning ($\nitrogen(p,\gamma)$ is the
slowest step in the CNO cycle), and this \nitrogen\ is converted to
\neon[22] during core helium burning. Since \neon[22] is neutron-rich,
its presence during the burning in the Type Ia supernova favors the
production of stable, neutron-rich nuclides like \nickel[58] and
\iron[54] at the expense of radioactive \nickel[56], the decay of
which powers the supernova.  Thus, more metal-rich progenitors are
expected to produce dimmer SNe~Ia.

The progenitor star's abundance of \iron[56] is also important, but
one possible limitation of TBT03 is that metallicity was treated as a
single variable, i.e. TBT03 make the simplifying assumption that O/Fe
is constant relative to Fe/H. In reality O/Fe varies as a function of
Fe/H (with a great deal of scatter), at least in the solar
neighborhood \citep[e.~g.][]{1989ARA&A..27..279W}, and the
degree of variation is expected to vary with the history of star
formation. Here we refine the estimate of the mass of \nickel[56] to
include changes in the relative abundances of O and Fe.

TBT03 showed that $M_{56}$, the mass of \nickel[56] synthesized in the
explosion, depended on the electron abundance \Ye,
\begin{equation}\label{e.base}
M_{56} = M_{56}^{0}(58\Ye-28).
\end{equation}
Here $M_{56}^{0}$ is the mass of \nickel[56] synthesized at an
electron abundance $\Ye=1/2$, i.~e. for a pure C-O white dwarf. It
does not include material that has been neutronized by \emph{in situ}
electron captures at high densities. To a good approximation, we can
consider the white dwarf to consist of \carbon, \oxygen, \neon[22],
and \iron. The electron abundance is then
\begin{equation}\label{e.Ye-def}
\Ye = \frac{1}{2}\left(1 - 22\YNe - 56\YFe\right) + 10 \YNe+26\YFe,
\end{equation}
where $Y_{i} = X_{i}/A_{i}$ is the abundance of nuclide $i$, $X_{i}$
and $A_{i}$ being the mass fraction and atomic mass number,
respectively.  Since \neon[22] is derived from the aboriginal CNO
abundances of the progenitor main-sequence star, $\YNe = \YO + \YN
+\YC$. Nitrogen is a secondary element, and we therefore write $\YN =
\YO\left[1+(\mathrm{N/O})\right]$, with $(\mathrm{N/O})= 0.0316+
126(\mathrm{O/H})$ \citep{1993MNRAS.265..199V}. This formula includes
the secondary production of nitrogen at high metallicity and agrees
with derived abundance ratios from the SDSS
\citep{2006ApJ...652..257L}. Carbon is a primary element, but is
mostly produced in lower-mass stars, which tends to make the ratio
$(\mathrm{C/Fe})$ nearly constant at subsolar metallicities, at least
down to $[\mathrm{Fe/H}]\gtrsim -2$
\citep{2003MNRAS.339...63C,2006MNRAS.367.1181B}, and there is evidence
that $[\mathrm{C/Fe}]$ decreases slightly at super-solar metallicity in
the thin disk \citep{2006MNRAS.367.1181B}. We take
$(\mathrm{C/Fe})=8.7$, its solar value \citep{2005ASPC..336...25A}.

Writing $\YO = X_{\mathrm{H}}(\mathrm{O/H})$ and $\YFe =
X_{\mathrm{H}}(\mathrm{Fe/H})$, setting $X_{\mathrm{H}} = 0.7392$
\citep{2005ASPC..336...25A}, and inserting equation~(\ref{e.Ye-def})
into equation~(\ref{e.base}) gives
\begin{equation}\label{e.m56}
\frac{M_{56}}{M_{56}^{0}} = 1 - 58\left\{ 0.763\left(\mathrm{\frac{O}{H}}\right) + 93.1 \left(\mathrm{\frac{O}{H}}\right)^{2} + 7.92\left(\mathrm{\frac{Fe}{H}}\right)\right\}.
\end{equation}
The factor of 58 comes from the assumption that the two
dominant nuclides in the burn to nuclear statistical equilibrium are
\nickel[56] and \nickel[58].  In actuality \iron[54] is also present,
and the ratio of \iron[54] to \nickel[58] depends on the thermal
conditions of the explosion (TBT03), so that the coefficient varies
between 58 and 54.  As shown by TBT03, this results is a small scatter
($<1\%$ at $Z = 3Z_{\sun}$) about the relation in eq.~(\ref{e.m56}).

\begin{figure}
\plotone{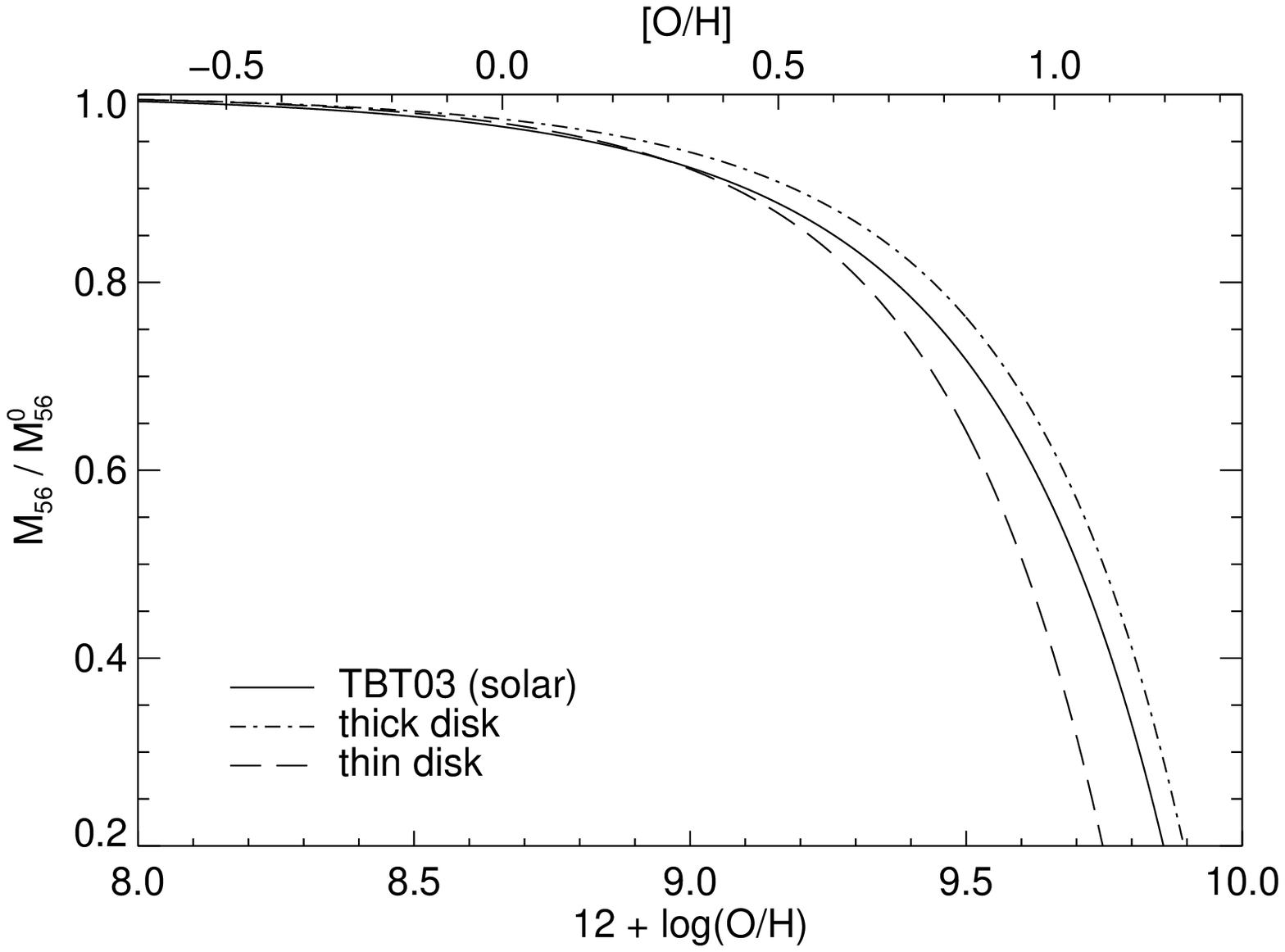} 
\caption{Theoretically expected
  variation in $M_{56}$, the mass of \nickel[56] synthesized, as a
  function of $\mathrm{O/H}$.  The solid line shows the prediction of
  TBT03, which fixed $[\mathrm{O/Fe}]=0$.  The other lines show the
  prediction when $\mathrm{O/Fe}$ varies according to linear relations
  \citet{2007A&A...465..271R} fitted to the thin (\emph{dashed lines})
  and thick (\emph{dot-dashed lines}) disk.  For illustrative
  purposes, we show each over a two decade span in metallicity as
  plotted in TBT03, though the thick and thin disk relations were
  derived over narrower metallicity ranges (see Table
  \ref{t.coefficients}), and SN hosts do not have average
  metallicities spanning this full range (section \ref{results}).}
\label{theory}
\vskip 0.2 in
\end{figure}

The two abundance ratios in equation~(\ref{e.m56}) are correlated,
albeit with a great deal of scatter.  We can therefore eliminate
$\mathrm{Fe/H}$ in favor of $\mathrm{O/H}$ and the coefficients from
the relation between \FeH\ (the ``metallicity'') and \OFe.  Here we
are using the standard notation $[\mathrm{A/B}] \equiv
\log(\mathrm{A}/\mathrm{B})-\log(\mathrm{A}/\mathrm{B})_{\odot}$.
\citet{2007A&A...465..271R} fit a set of linear relations,
$\OFe = a + b\FeH$, for the thin disk, the thick disk, and the halo
(see Table~\ref{t.coefficients}). We therefore write $\mathrm{Fe/H} =
\left(\mathrm{Fe/H}\right)_{\odot}10^{\FeH}$, use the linear relation
between \OFe\ and \FeH\ and the identity $\OFe = \OH-\FeH$ to obtain
$\FeH = (\OH-a)/(1+b)$, and insert the solar photospheric abundances
\citep{2005ASPC..336...25A},
$\log\left(\mathrm{O/H}\right)_{\odot} = -3.34$ and
$\log\left(\mathrm{Fe/H}\right)_{\odot} = -4.55$, into
equation~(\ref{e.m56}) to obtain
\begin{equation}\label{e.m56-oh}
\frac{M_{56}}{M_{56}^{0}} = 1 - 0.044\left[\frac{(\mathrm{O/H})}{10^{-3}}\right]
        \left\{1 + 0.122\left[\frac{(\mathrm{O/H})}{10^{-3}}\right] +
        10^{-(0.19+0.53b+a)/(1+b)}\left[\frac{(\mathrm{O/H})}{10^{-3}}\right]^{-b/(1+b)}\right\}.
\end{equation}
For many purposes, it is more convenient to work directly with \FeH\ 
as the primary variable.  We therefore write $(\mathrm{O/H}) =
(\mathrm{O/H})_{\odot} 10^{\OFe+\FeH}$ and use the linear relation
between \OFe\ and \FeH\ to obtain
\begin{equation}\label{e.m56-feh}
\frac{M_{56}}{M_{56}^{0}} = 1 - 0.020\times 10^{a+(1+b)\FeH}
        \left\{ 1 + 0.056\times 10^{a+(1+b)\FeH} + 0.64\times 10^{-a-b\FeH}\right\}.
\end{equation}
In both equations, (\ref{e.m56-oh}) and (\ref{e.m56-feh}), setting
$a=b=0$ will give the case of an system where $\mathrm{O/Fe}$ is held
at its solar value.  Figure~\ref{theory} shows the
equation~(\ref{e.m56-oh}) with coefficients taken from
\citet{2007A&A...465..271R} for the thin disk (\emph{dashed
  line}) and thick disk (\emph{dash-dotted line}). We also show, for
comparison, the relation when $\mathrm{O/Fe}$ is fixed at its solar
value (\emph{solid line}).  This last curve differs slightly from that
in TBT03 because we are using the \citet{2005ASPC..336...25A}
abundances for the solar composition, and we include the secondary
production of \nitrogen.  This scaling is appropriate for
the Milky Way, but supernova hosts may have a different metallicity
history. This does not appear to be much of a concern, however, since
the differences between the predictions using the thick and thin disk
parameters are small for $\OH < 0.5$. We conclude from this section
that O/Fe variations can safely be neglected at the level of precision
of the TBT03 theory, for the relevant range, $\OH < 0.5$.
We note from Figure~\ref{theory} that TBT03 would have predicted a
variation of nickel mass with metallicity of less than 10\%
had they restricted their considerations to the observationally
relevant range of $\OH < 0.5$ that corresponds to $\FeH < 0.6$
in the thin disk model.

\begin{deluxetable}{lccc}

\tablewidth{0pt}
\tablecaption{$\OFe$ for different populations}
\tablehead{ \colhead{} & \colhead{a} & \colhead{b} & \colhead{range}} 
\startdata
Thin disk & $0.096\pm 0.004$ & $-0.327\pm0.016$ & $-1.0< \rm [Fe/H] < -0.3$\\
Thick disk & $0.370\pm0.027$ & $-0.121\pm0.043$ & $-0.8 < \rm [Fe/H] < 0.3$\\
Halo & $0.388\pm0.049$ & $-0.048\pm0.071$ & $-1.4 < \rm [Fe/H] < -0.4$
\enddata
\tablecomments{Coefficients of the fit $\OFe = a+b\FeH$ from \protect\citet{2007A&A...465..271R}.\label{t.coefficients}}
\end{deluxetable}

\section{Method}
\label{method}
\subsection{Estimating host galaxy masses}
To test the TBT03 theory, we must determine host galaxy
metallicities, and to do that we first determine galaxy masses
from model SED fits to SNLS photometry.  Following the procedure
outlined in \citet{2006ApJ...648..868S}, we used the code Z-PEG
\citep{2002A&A...386..446L} to fit PEGASE.2
\citep{1997A&A...326..950F,1999astro.ph.9912179} models to host galaxy
\ugriz\ photometry from the SNLS to determine their masses and star
formation rates.  We extend the analysis of
\citet{2006ApJ...648..868S} to include data from the SNLS 3rd year
cosomology sample \citep{2008sullivan,2008conley,2008guy}.  

The SNLS revisits 4 one-degree fields five times per
lunation for the 5-6 lunations per year that each field is visible.
Images with seeing $< 0.8 \arcsec $ and no SN light are combined to
provide a deep, multi-year reference image from which the host galaxy
properties are measured.  SN types and host galaxy redshifts were determined from Gemini, VLT, and Keck spectroscopy \citep{2005ApJ...634.1190H,2008A&A...477..717B,2008arXiv0809.4407B,2008ApJ...674...51E}.
  We categorize each galaxy by its star
formation rate per unit mass (specific star formation rate; sSFR) into
strong star formers ($\rm log (sSFR) > -9.5$), weak star formers ($\rm
-12 > log (sSFR) > -9.5$), and passive galaxies with no measurable
star formation rate ($\rm log (sSFR) < -12$).  Here, SFR is the
average star formation rate over the last 0.5 Gyr, which is estimated from the optical SED
fitting.  We assign errors
corresponding to the range of masses that can be fit within a model
given the photometric errors, or the scatter between acceptable model
fits in terms of $\chi^2$, whichever is larger.  As in
\citet{2006ApJ...648..868S}, we only use SNe at $z\leq 0.75$ where the
SNLS sample is relatively complete.  For more information on the SNLS,
including discoveries, the photometric system, and SN typing, see
\citet{2006A&A...447...31A}.  

\subsection{Estimating metallicity}
To determine average host galaxy metallicities we use the
\citet{2004ApJ...613..898T} correlation between host galaxy stellar
mass and gas phase O/H as determined from 53,000 galaxies in the Sloan
Digital Sky Survey.  There is a 0.1 dex scatter in this relationship.
For galaxies with log $M_* < 8.4$ we use the low mass extension of
\citet{2006ApJ...647..970L}.  Both of these relationships were derived
for zero redshift --- \citet{2005ApJ...635..260S} find that
higher-redshift galaxies have lower average gas phase metallicities.  They give
an ``effective Mass'' versus redshift relation
(\citet{2005ApJ...635..260S} equation 9), which we use here to
adapt the \citet{2004ApJ...613..898T} and \citet{2006ApJ...647..970L}
relations to the redshifts of our galaxies.  

Technically the \citet{2004ApJ...613..898T} relationship only covers
gas phase metallicity, and is not guaranteed to be applicable in
elliptical galaxies, where there is generally little cold gas.  Still,
there are several reasons not to dismiss the sample of passive
galaxies from this analysis.  First, the amount of star formation in
these galaxies is inferred from their broadband colors, not from
spectroscopy or morphology, so the presence of gas is not ruled out.
Second, because the mechanism for retaining metals is the depth of the
galaxy gravitational potential well, one might still expect
ellipticals to follow this trend, independent of the presence of gas.
The SN progenitor was born when there was star formation, and
thus much more gas in the galaxy.  Finally, we caution that stellar
metallicity and gas phase metallicity are not identical, though they
are correlated \citep[e.g.][]{2005MNRAS.358..363C}.

\subsection{Estimating \nickel[56] mass}
The mass of \nickel[56] can be estimated from the peak bolometric
luminosity and the rise time of a SN Ia using Arnett's Rule
\citep{1979ApJ...230L..37A,1982ApJ...253..785A}, that the luminosity
radiated at maximum light is approximately equal to the energy
deposited by radioactive decay:
\begin{displaymath}
M_{\rm Ni}=\frac{L_{\rm bol}}{\gamma \dot{S}(t_R)},
\end{displaymath}
where $\gamma$ is the ratio of bolometric to radioactivity
luminosities, near unity
\citep[e.g.][]{1992ApJ...392...35B,1996ApJ...457..500H,2006A&A...453..229B}.
Here we adopt a value of $1.2\pm0.1$
\citep{1995PhRvL..75..394N,1995PhR...256...53B} to be consistent with
previous SNLS work \citep{2006Natur.443..308H}, though a global change
to this number does not affect our conclusions.  $\dot{S}$ is the
radioactivity luminosity per solar mass of $^{56}$Ni from its decay to
$^{56}$Co and subsequent decay to $^{56}$Fe:
\begin{displaymath}
\dot{S} = 7.74 \times 10^{43} e^{-t_R/8.8} +  1.43 \times 10^{43}
[e^{-t_R/111} - e^{-t_R/8.8}]\,\, \rm 
erg\,\, s^{-1}\,\, M_{\odot}^{-1}.
\end{displaymath}

Here $t_R$ is the time in days for the supernova to rise from
explosion to maximum light: $t_R=s \times 19.5$ days
\citep{1999AJ....118.2675R,2006AJ....132.1707C}, where $s$ is the
``stretch'', a scaling factor for the time axis of the SN lightcurve
that correlates with luminosity
\citep{1997ApJ...483..565P,2008arXiv0803.3441C}.  Note that simply
taking the mass difference between \nickel[56] and \cobalt[56], as
some authors have done, will overestimate the energy production.  The
heating from the decay of \nickel[56] is primarily from the radiative decay
of the 1.7 MeV nuclear level in \cobalt[56] \citep{1999NDS....86..315J}. We include
contributions from the positron decay branching in the heating from
the decay of \cobalt[56] \citep{Colgate1980The-Luminosity-}.

The least straightforward part of determining \nickel[56] mass from
SN Ia lightcurves is estimating the bolometric luminosity from incomplete
data.  For rare, very well-observed
supernovae with data spanning much of the UVOIR (ultraviolet, optical,
infrared) region, all of the flux can be added, and only a small
correction is necessary for missing data
\citep[e.g.][]{2000A&A...359..876C,2006A&A...460..793S}.   However, this has to date only
been possible for small numbers ($\sim 15$) of low-redshift supernovae.  Because
the best-observed supernovae tend to be the closest and brightest, and
thus not in the Hubble flow, these
SNe often have uncertain distances and luminosities. 

Other authors have used the luminosity in a single band, combined with
a bolometric correction, to estimate the bolometric flux
\citep[e.g.][]{1992ApJ...392...35B, 2006Natur.443..308H}.  This allows
the study of greater numbers of supernovae, achieving enhanced global
accuracy, at the expense of individual accuracy.  It also makes
possible the study of high redshift supernovae where the restframe NIR
is inaccessible.

\begin{figure}
\plotone{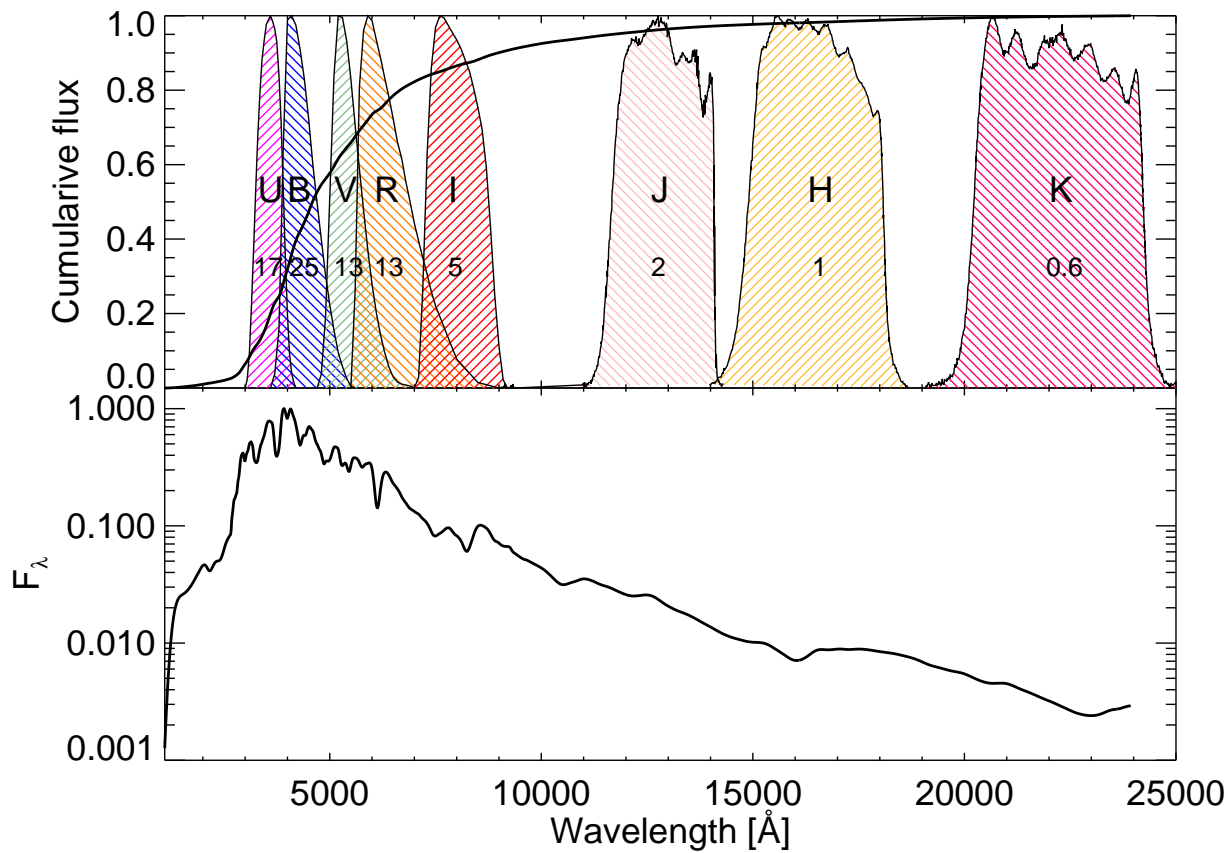} 
\caption{{\it Bottom:} The \citet{2007ApJ...663.1187H} SN Ia maximum-light spectral energy
  distribution template (plotted on a log axis), updated to include
  NIR data. {\it Top:} Cumulative
  flux as a function of wavelength.  Johnson-Cousins filters are shown
  with labels corresponding to the percentage of total flux through
  each. Approximately 3/4 of the flux for a typical SN Ia is radiated
  in the optical.\label{cumflux}}
\end{figure}

Bolometric corrections work well at maximum light for Type Ia
supernovae, because the peak of
emission is always in the restframe $B$ band, near 4000\AA\ (in the
absence of strong reddening).
Figure~\ref{cumflux} (bottom panel) shows a typical SN Ia spectral
energy distribution at maximum light (from
\citet{2007ApJ...663.1187H}, updated with NIR data).  The upper panel
shows the cumulative flux as a function of wavelength.  Only $\sim
5\%$ of the SN Ia flux is emitted in the UV shorward of the $U$
filter, and less than 10\% is emitted in the IR.\footnote{Our template
  spans 1000\AA --- 2.4 $\mu$m -- since there is no evidence for
  significant SN Ia flux outside of this range,  we assume it to be
  negligible.}  Since roughly one-quarter of the SN Ia flux is emitted
in the $B$ band, and this fraction is relatively constant from SN to
SN, it is easy to see why the $B$ band is the most commonly used
indicator of SN Ia luminosity.

\begin{figure}
\plotone{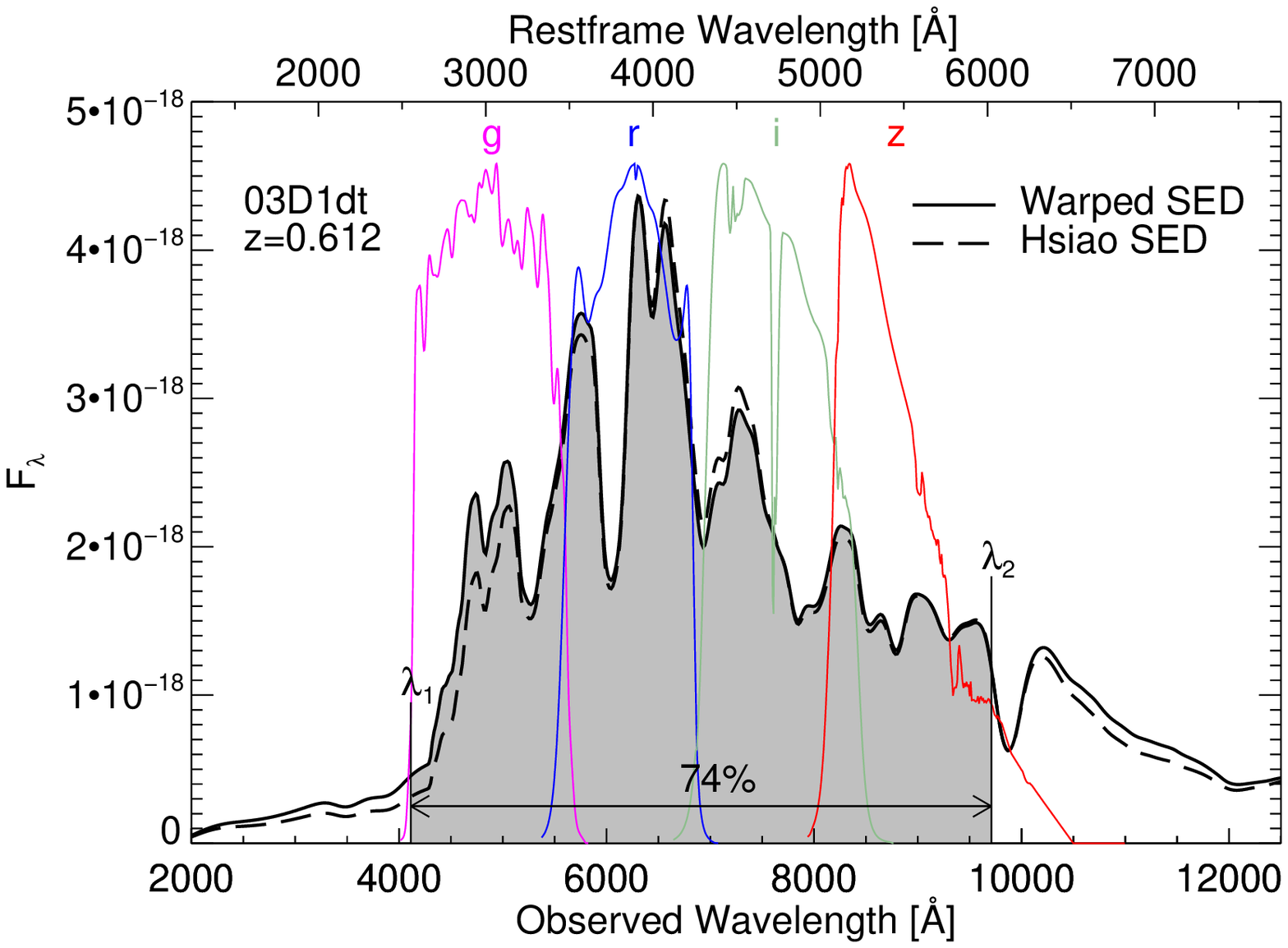} 
\caption{Determination of the bolometric flux for an example SN
  Ia, SNLS-03D1dt at $z=0.612$.  The dashed line shows the
  \citet{2007ApJ...663.1187H} SED, and the solid line shows it warped
  to match the observed \griz\ fluxes.  This warped SED is then
  integrated over the observed region (shaded), from $\lambda_1$ to
  $\lambda_2$, arbitrarily defined to start and stop where the bluest
  and reddest filters are 1/5 of their maximum height.  Since this
  region contains only 74\% of the total Hsiao SED bolometric flux
  (see Fig.~\ref{cumflux}), the integrated flux is divided by 0.74 to
  obtain the bolometric flux.\label{bolint}}
\end{figure}

Here we adopt a new approach to bolometric flux estimation that
exploits the excellent multiwavelength coverage of SNLS supernovae.
For each SNLS SN Ia, we smoothly adjust (i.e. ``warp'') the Hsiao SED
template so that it has the observed fluxes in \griz .  We then
integrate the template SED over the wavelength interval where we have
observations, and correct for the unobserved fraction.  We turn this
bolometric observed flux into a luminosity using the redshift and a
flat cosmology with $H_0$=70 km s$^{-1}$ Mpc$^{-1}$ and
$\Omega_M=0.3$.  An example is shown in Figure~\ref{bolint}.  

Note that this process uses much of the same machinery as the 
SiFTO lightcurve fitter \citep{2008arXiv0803.3441C}.  The first step,
warping the SED to match the observed colors, is identical.  But in
the next step the methods diverge -- SiFTO calculates the restframe
flux of SN through another filter, whereas here we calculate the
integrated flux over the region spanning all observed filters.

Mathematically, we define $f$ as the fraction of the bolometric flux
covered by our observations, such that:
\begin{equation}\label{fraction}
f=\frac{\int_{\lambda_1}^{\lambda_2} H(\lambda) d\lambda}{\int
  H(\lambda) d\lambda} 
\end{equation}
where $H(\lambda )$ is the Hsiao SN Ia SED
template, $\lambda_1$ is the wavelength at 1/5 the filter height of
the bluest filter for which there is data, and $\lambda_2$ is the
wavelength at 1/5 the filter height of the reddest filter for which
there is data.  The choice of 1/5 of the filter height as the cutoff
is arbitrary -- changing it does not affect our results.

We define $S(\lambda )$ as the SED after it is warped to match the
observed colors, so that the bolometric flux is 
\begin{equation}\label{bolflux}
F_{bol}=\frac{\int_{\lambda_1}^{\lambda_2} S(\lambda) d\lambda}{f}.
\end{equation}  Flux errors are given by the SiFTO covariance matrix
\citep[see][]{2008arXiv0803.3441C}.  Errors on the other quantities are:
\begin{equation}\label{sigsdot}
\sigma_{\Sdot}=|-1.4\times 10^{44} e^{-2.2 s}-2.5\times 10^{42} e^{-0.18s}| \times \sigma_s
\end{equation}
\begin{equation}\label{signi}
\sigma_{\nickel[56]}=\sqrt{(\frac{1}{\gamma \Sdot})^2\sigma_{L_{bol}}^2+(\frac{L_{bol}}{\gamma \Sdot^2})^2 \sigma^2_{\Sdot}+(\frac{L_{bol}}{\gamma^2 \Sdot})^2 \sigma^2_{\gamma}}.
\end{equation}
 Based on constraints from theory
\citep[e.g.][]{1992ApJ...392...35B, 1996ApJ...457..500H}, and from
estimating \nickel[56] using different empirical methods
\citep{2006A&A...460..793S}, we adopt $\sigma_{\gamma}=0.1$ and
generally this term dominates the error budget.

\begin{deluxetable}{lllll}
  \tablewidth{3.1in} \tablecaption{Parameters of color cut}
  \tablehead{\colhead{} & \colhead{color} & \colhead{$a$} & \colhead{$b$}&
    \colhead{$c$}} \startdata
  $U02-B$ & 0.1856& -0.0003& -0.1681&  0.0790\\
  $U-B$  & 0.4031&      0.1816&     -0.0567&     0.0477\\
  $V-R$ & 1.0129& 0.1957& -0.5634& 0.0646 \enddata \tablecomments{SNe
    Ia rejected if the $B-V$ color is redder than $a({\rm
      color}+0.5) + b(s-1)+c$, where $s$ is the stretch (lightcurve
    width).  ``$U02$'' is a virtual filter useful in k-correcting
    \citep{2007ApJ...663.1187H} high redshift data, defined to be
    bluer than the $U$ filter \citep{2008sullivan}.  A $B-V$ and at
    least one other color was required for all SNe Ia.  Cuts were made
    in as many filters as data were available.  The cuts mimic those
    used in \citep{2008sullivan}.\label{ccut}}
\end{deluxetable}

In the absence of dust extinction, one should use the raw absolute
magnitude (i.e. not corrected for stretch 
or color), because
this reflects the true luminosity of the SN.  The complication is that
the observed color is a mixture of reddening due to dust (which should be
corrected for), and intrinsic SN color (which should not be
corrected).  If the color correction were primarily due to normal
dust, it would be expected to follow the Milky Way extinction law
\citep{1989ApJ...345..245C}, but it does not
\citep[e.g.][]{2008ApJ...675..626W,2008MNRAS.384..107E,2007AJ....133...58K,2007ApJ...664L..13C}.
There are theoretical reasons to expect, and some evidence, that the
majority of SNe Ia suffer little extinction \citep{1998ApJ...502..177H,
  1999ApJ...517..565P, 1999AJ....118.1766P, 2004NewAR..48..567C},
though some of the reddest outliers surely have significant dust
extinction.  As our primary result, we chose to make a color cut
\citep{2008sullivan} to eliminate the most extincted SNe rather than
making an uncertain extinction correction \citep{2007ApJ...664L..13C}.
The parameters of the color cut are given in Table~\ref{ccut}.  The
chosen cut parameters, and whether or not the cuts are made at all, do
not affect the conclusions.  We also study the effect of making a
color correction, which corrects for dust, but also improperly
corrects intrinsic luminosity dispersion.  Such results give a lower
limit on the dispersion of $^{56}$Ni.

To make the color correction we followed the same steps as above, but first
we use the color excess measured for each SN Ia to unredden the \citet{2007ApJ...663.1187H} SED template using
the \citet{1989ApJ...345..245C} law with $R_V=1.8$
\citep{2007ApJ...664L..13C}.  We use a fiducial color of $c=0$, where
$c$ is a linear combination of restframe $B-V$ and $U-B$
\citep{2008arXiv0803.3441C}, roughly corresponding to 
$E(B-V)_0=-0.057$ \citep{2006A&A...447...31A}.

Mathematically, if $H_0(\lambda)$ is the unreddened template, then
this process can be summed up by the following two equations: 
\begin{equation}\label{fract2}
f\arcmin=\frac{\int_{\lambda_1}^{\lambda_2} H(\lambda) d\lambda}{\int
  H_0(\lambda) d\lambda}, 
\end{equation}
and
\begin{equation}\label{bolflux2}
F_{bol}=\frac{\int_{\lambda_1}^{\lambda_2} S(\lambda) d\lambda}{f\arcmin}.
\end{equation}  

\section{Results}
\label{results}
\subsection{Galaxy Mass vs. SN Ia stretch}\label{stretchmass}
\begin{figure} 
  \plotone{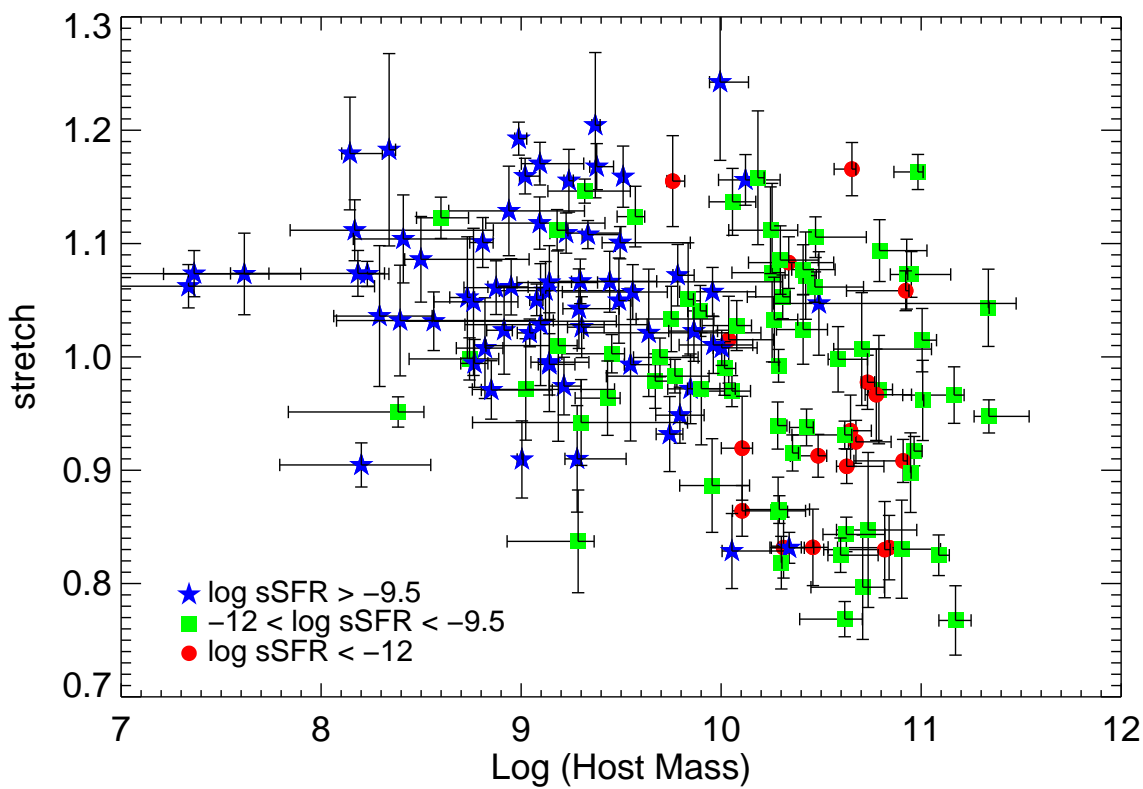} 
\caption{SN relative lightcurve width (stretch) as a function of host
  galaxy mass determined from fitting PEGASE.2 galaxy models to host galaxy
  \ugriz\ photometry.  Circles (red online) have hosts with no detectable star
  formation (log $\rm sSFR < -12$), where sSFR is specific star
  formation rate.  Star symbols (blue) denote strong star formation (log sSFR$
  >-9.5$), and squares (green) are in between.  \label{massstr}}
\end{figure}

It is well known that SN lightcurve widths correlate with host
galaxy morphology \citep{1995AJ....109....1H, 2000AJ....120.1479H,
  2001ApJ...554L.193H, 2005ApJ...634..210G}, host color
\citep{1996ApJ...465...73B}, or specific star formation rate
\citep{2006ApJ...648..868S}, in the sense that SNe with wider
lightcurves are more commonly associated with late-type galaxies and vice
versa.  The fact that elliptical galaxies are generally larger (in a
spatial sense) than spirals also may account for trends seen between
lightcurve width and projected galactocentric distance
\citep{1997ApJ...483L..29W,2005ApJ...634..210G}.  And while SN rates
have been studied as a function of host galaxy mass
\citep{2005A&A...433..807M, 2006ApJ...648..868S}, no study has
examined lightcurve width as a function of host galaxy mass.  Figure
\ref{massstr} shows such a comparison for SNLS SNe Ia.  Keeping in
mind that passive galaxies tend to have a high mass, 
the results are as expected from previous work
--- high mass galaxies host few high stretch SNe Ia, and low mass
galaxies host few low-stretch SNe Ia.

\begin{figure} 
  \plotone{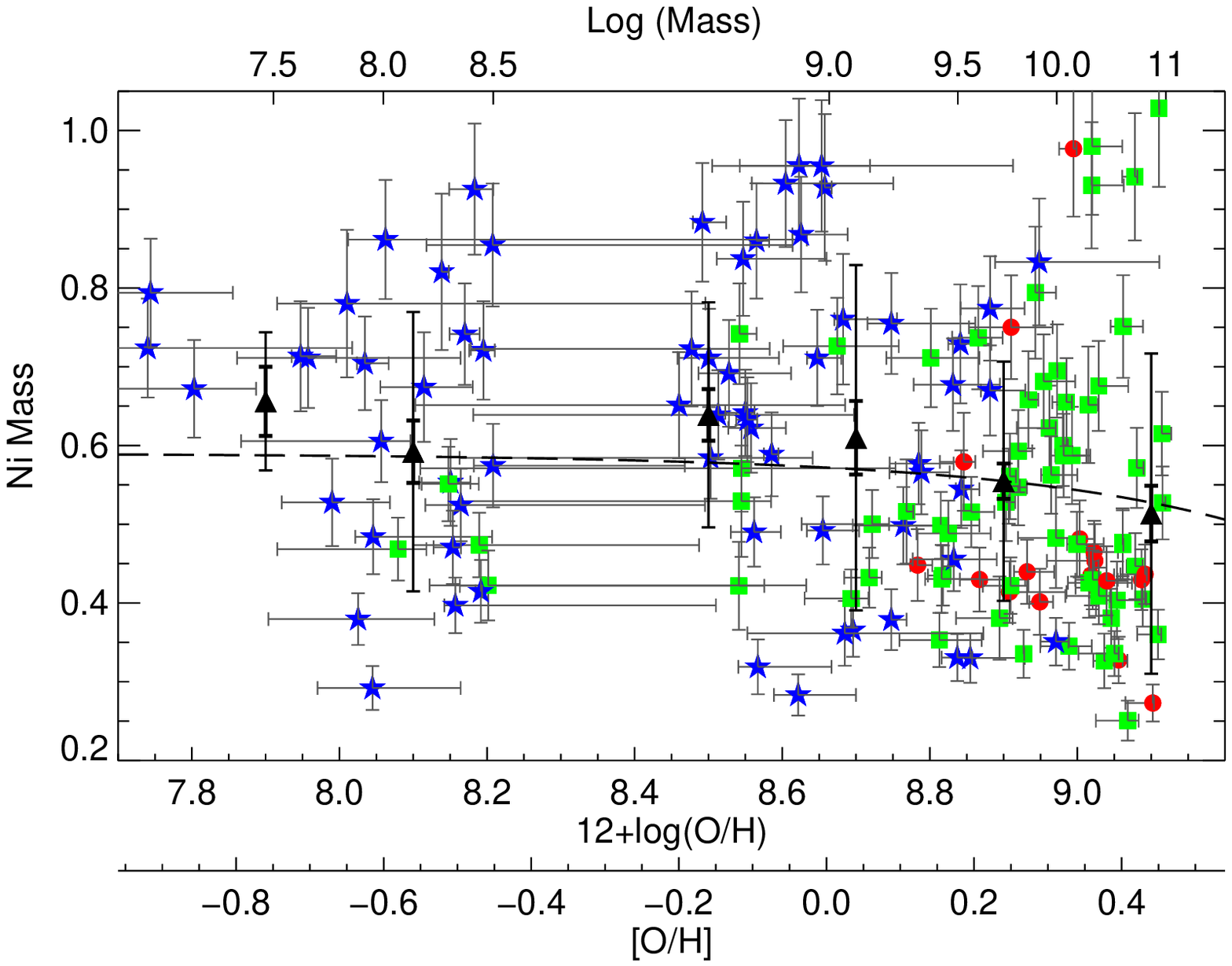} 
\caption{\nickel[56] mass as a function of inferred average host gas phase metallicity.
  The dashed line is the prediction of TBT03 modified to use the thin
  disk variation in O/Fe as described in \S \ref{theorymod}.
  Points are SNLS SNe Ia with \nickel[56] mass and inferred average host
  metallicity determined as described in the text and colored as in
  Fig.~\ref{massstr}.  Vertical error bars include flux errors,
  stretch errors, and errors associated
  with Arnett's Rule.  Horizontal error bars include errors from the
  model fits, but for clarity do not include the 0.1 dex scatter from the
  \citet{2004ApJ...613..898T} relationship.  The gap at
  12+log(O/H)=8.3 arises from the discontinuity between the
  \citet{2004ApJ...613..898T} mass-metallicity relationship and the
  \citet{2006ApJ...647..970L} low mass extension (there is no gap in
  host galaxy masses as can be seen from Fig.~\ref{massstr}).  Points are
  plotted by metallicity, and are corrected for redshift
  mass-metallicity effects as mentioned in the text, but the upper axis gives an approximate
  corresponding mass at zero redshift. Triangles are averages in 0.2
  dex bins.  Outer error bars are the standard deviation of the points
  in that bin, and the inner error bars are the errors on the mean.\label{metal}}
\end{figure}

\subsection{\nickel[56] Mass vs host galaxy metallicity}\label{massmetal}
In Figure~\ref{metal}, we have converted galaxy mass to gas phase
metallicity and converted integrated flux to $^{56}$Ni mass as
outlined in \S \ref{method}.  We have overplotted the TBT03
relationship between metallicity and predicted SN Ia \nickel[56]
yield, corrected for O/Fe differences as outlined in \S
\ref{theorymod}.  In this figure we make a color cut, but not a color
correction.  Figure~\ref{metal} shows that there is a drop in average
\nickel[56] mass for SNe Ia from high mass, high metallicity
(12+log(O/H)$>$ 8.8) galaxies.  TBT03 predicted a 25\% difference in
\nickel[56] yield over a factor of 3 difference in progenitor
metallicity.  However, most of the effect occurred at metallicities
12+log(O/H) $>$ 9.2 (see Fig.~\ref{theory}), whereas most SNe Ia in
this study occur in galaxies with lower metallicity.
Figure~\ref{metal} shows that the TBT03 prediction is fairly flat over
the range of metallicities determined for actual SN hosts.  Even when
the steeper ``thin disk'' relation is used from Fig.~\ref{theory}, at
most, the expected effect is only 0.06 \Msun\ in this range.  In
comparison SNe Ia show a wide range in derived \nickel[56] mass, here
0.2 to 1.0 \Msun , though ranging from 0.1 to 1.3 \Msun\ for extreme
cases like SN~1991bg \citep{1992AJ....104.1543F} and SNLS-03D3bb
\citep{2006Natur.443..308H}.

In other words, the TBT03 theory appears to be qualitatively
consistent with observations, though there is additional scatter not
explained by the theory.  An additional error of 0.16 \Msun\ of
\nickel[56] had to be added to the \nickel[56] error bars in
quadrature to achieve $\chi^2/\mathrm{DOF}=1$.

Several SNe Ia with \nickel[56] below 0.2 \Msun\ were
conservatively eliminated with the color cut, on the grounds that
extinction might be causing a low derived \nickel[56] mass, although
is also possible that these are subluminous SNe, which have
intrinsically redder colors \citep{2004ApJ...613.1120G}.  

Various effects can systematically alter the scaling of the observed
\nickel[56] mass in Fig. \ref{metal}, including changing the Hubble
constant $H_0$ or the ratio of radiated to deposited radioactive decay
energy $\gamma$, but these
do not affect the overall results, because the normalization of the
TBT03 prediction is arbitrary.  Here we have chosen the normalization
by fitting to the data.

\begin{figure} 
  \plotone{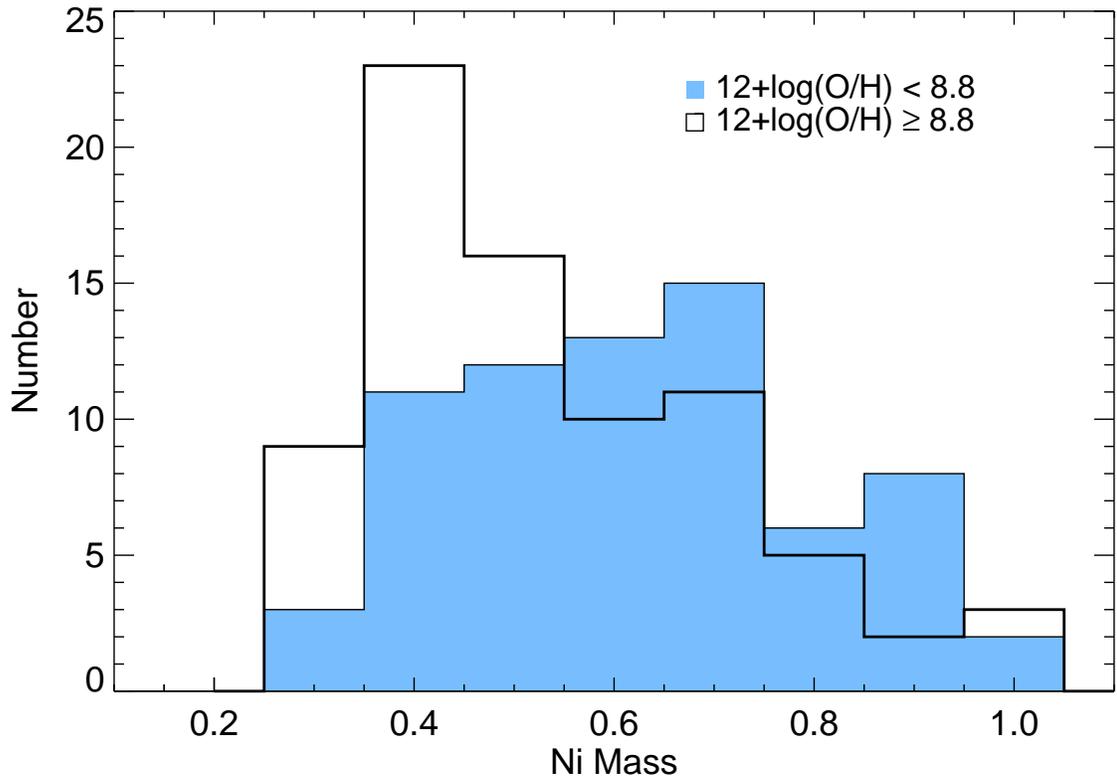} 
\caption{Histogram of SN $^{56}$Ni mass split by inferred host gas phase metallicity.
  Higher metallicity hosts have a lower \nickel[56] yield, on
  average.  This uses the non-color-corrected data from
  Fig.~\ref{metal}.
\label{metalhist}}
\end{figure}

Another way to visualize the difference in \nickel[56] yield between
high and low metallicity galaxies is the histogram in Figure~\ref{metalhist}.
Supernovae in more metal-rich hosts (12+log(O/H)$>$ 8.8) have a much lower
average \nickel[56] yield than those in metal-poor hosts.  The high
metallicity set has a peak in the distribution near 0.4 \Msun , while
the low metallicity set peaks near 0.7 \Msun . The
probability that the \nickel[56] yields from high and low metallicity
hosts were drawn from the same distribution, according to the
Kolmogorov-Smirnov (KS) test is $9 \times 10^{-3}$.

\begin{figure} 
  \plotone{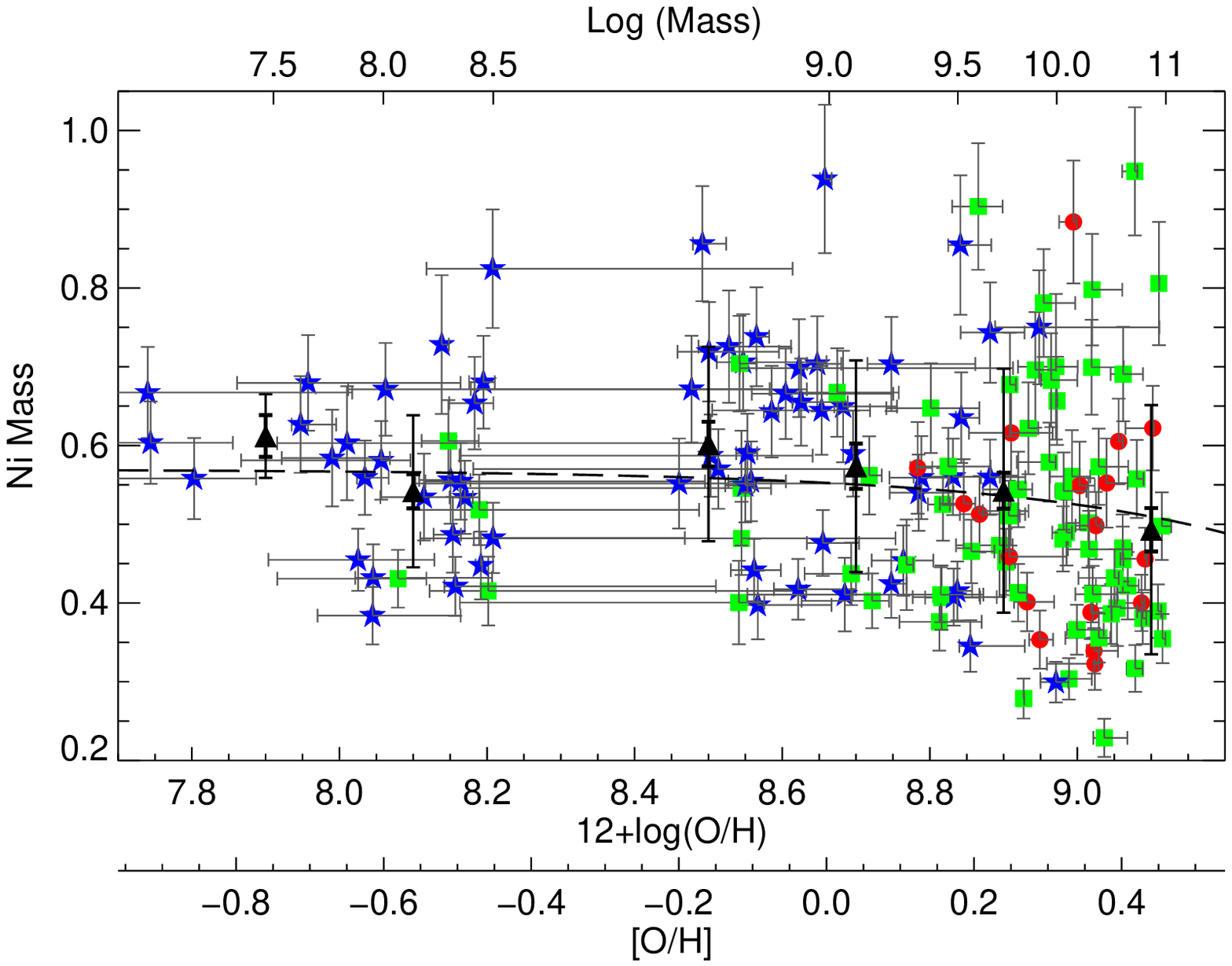} 
\caption{Same as Fig. \ref{metal}, but a color correction has been
  made.  This correction would be appropriate if all SNe Ia were the
  same color, and all reddening and dimness were due to dust.
  However, there is an intrinsic color-luminosity relationship for SNe
  Ia that it is not currently possible to separate distinctly.  Thus a
  color correction overcorrects the luminosity, and the result is a
  lower limit on the scatter in SN Ia $^{56}$Ni mass. 
\label{metal-colorcor}}
\end{figure}

\subsection{Reddening correction}\label{reddening}

It is possible that the above results overestimate the scatter in SN
Ia \nickel[56] mass, if significant dust extinction remains after the
color cut described in \S \ref{method}.  To test this, we make the
color correction in Figure~\ref{metal-colorcor}.  As expected, the
scatter is reduced -- now only 0.12 \Msun\ must be added in quadrature
to the error bars to give $\chi^2/\mathrm{DOF}=1$.
Since this color correction includes an intrinsic SN Ia
color correction as well as a dust correction, it will over-correct
the SN Ia magnitudes, but we do it to get an estimate of the limiting
case.  If SNe Ia are color-corrected using $\beta=2.8$
\citep{2007ApJ...664L..13C}, the scatter is reduced to a range of 0.6
\Msun\ in \nickel[56] for all but a few outliers.  Thus the upper
limit on the TBT03 metallicity effect is $0.06/0.6=10\%$.

\begin{figure} 
  \plotone{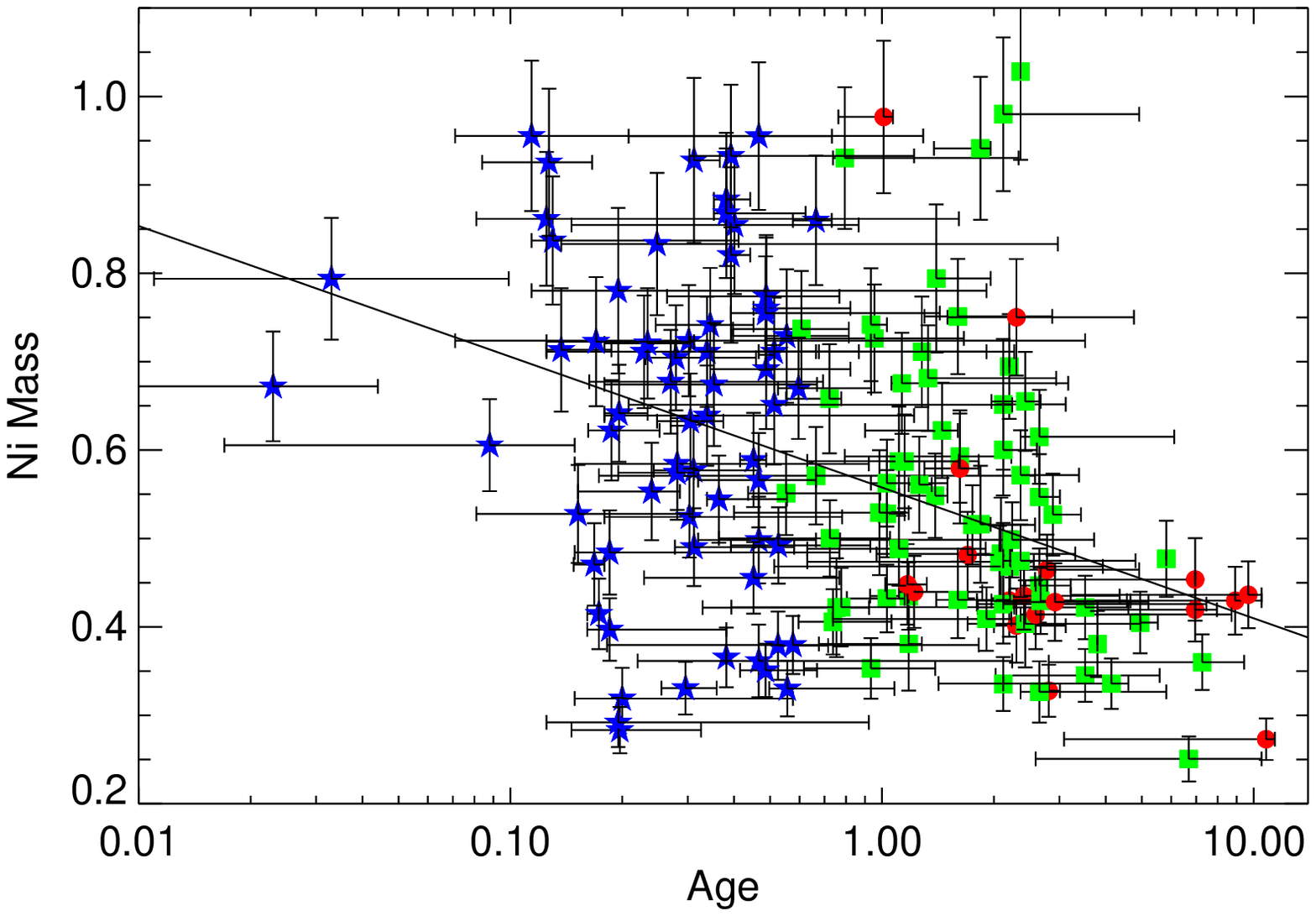} 
\caption{$^{56}$Ni mass versus host luminosity weighted age as
  estimated from PEGASE.2 models.  A color cut on the SN magnitudes has been made, but not a
  color correction.  The line shows a fit to the data with slope
  $-0.15\pm 0.03$ and correlation coefficient -0.38.\label{ageni}}
\end{figure}

\begin{figure} 
  \plotone{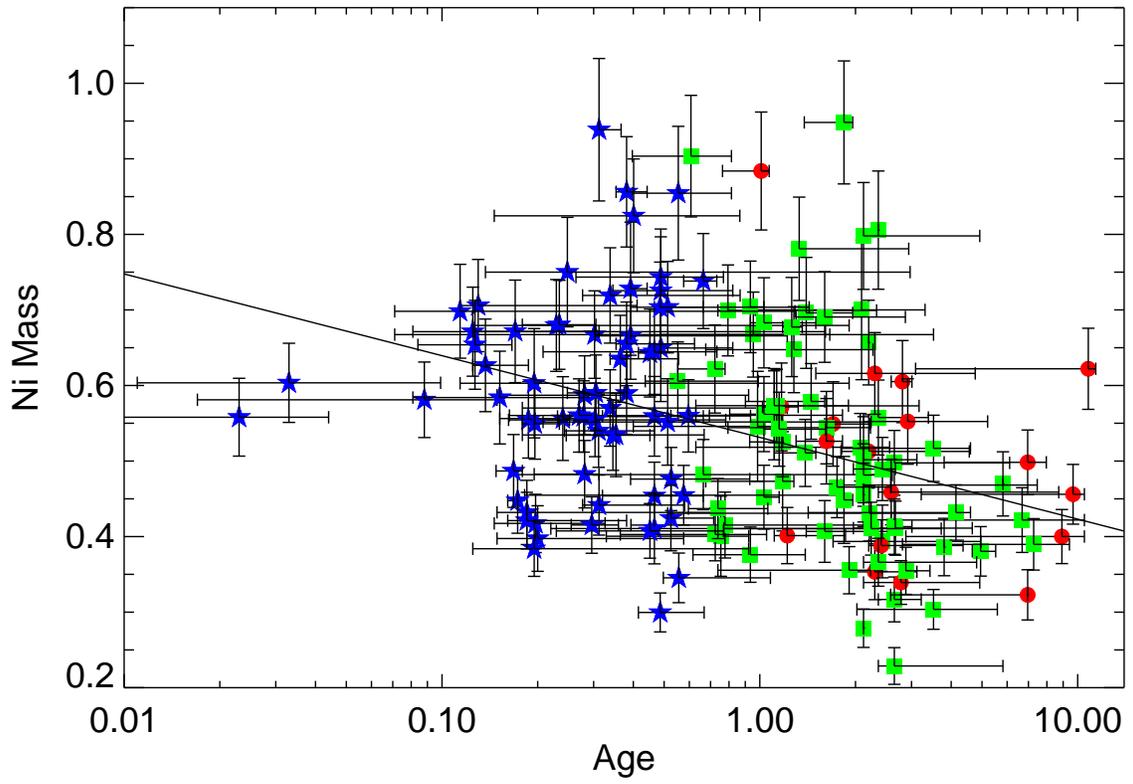} 
\caption{Same as Fig. \ref{ageni}, but a color correction has been
  made.   The line shows a fit to the data with slope
  $-0.11\pm 0.028$ and correlation coefficient -0.37.
\label{ageni-colorcor}}
\end{figure}


\subsection{Progenitor Age}\label{agesec}

Since there is an age-metallicity degeneracy, it is difficult to
separate effects that may be caused by metallicity from those that may
be caused by age of the progenitor.  An age-brightness effect must
exist for SNe Ia, since both SN rate and average luminosity are
increased in young stellar populations
\citep[e.g.][]{2001ApJ...554L.193H,2005A&A...433..807M,2006ApJ...648..868S}.
Figure~\ref{ageni} confirms that there is a correlation between SN
$^{56}$Ni mass and host luminosity-weighted age as estimated from
PEGASE.2 models --- the slope of the line is significant at $\sim
5\sigma$.  This correlation is understandably noisy -- the PEGASE.2
luminosity-weighted age is merely an indicator of the mean age of an
entire galaxy -- the SN progenitor may have a very different age.  
Any age-brightness relation must be related to
with the mass of the secondary star -- in an old population more
massive secondaries do not exist.  But this is difficult to understand
in the Chandrasekhar mass model -- if all SNe Ia explode when the
primary nears the Chandrasekhar mass, why should the mass of the
secondary matter?  This might be an indication that the binary
evolution and mass transfer history, which is strongly dependent on
the mass of the secondary, influences the outcome of the explosion.

\subsection{\nickel[56] mass vs. metallicity grouped by stretch}\label{ni3psec}

\begin{figure}
  \plotone{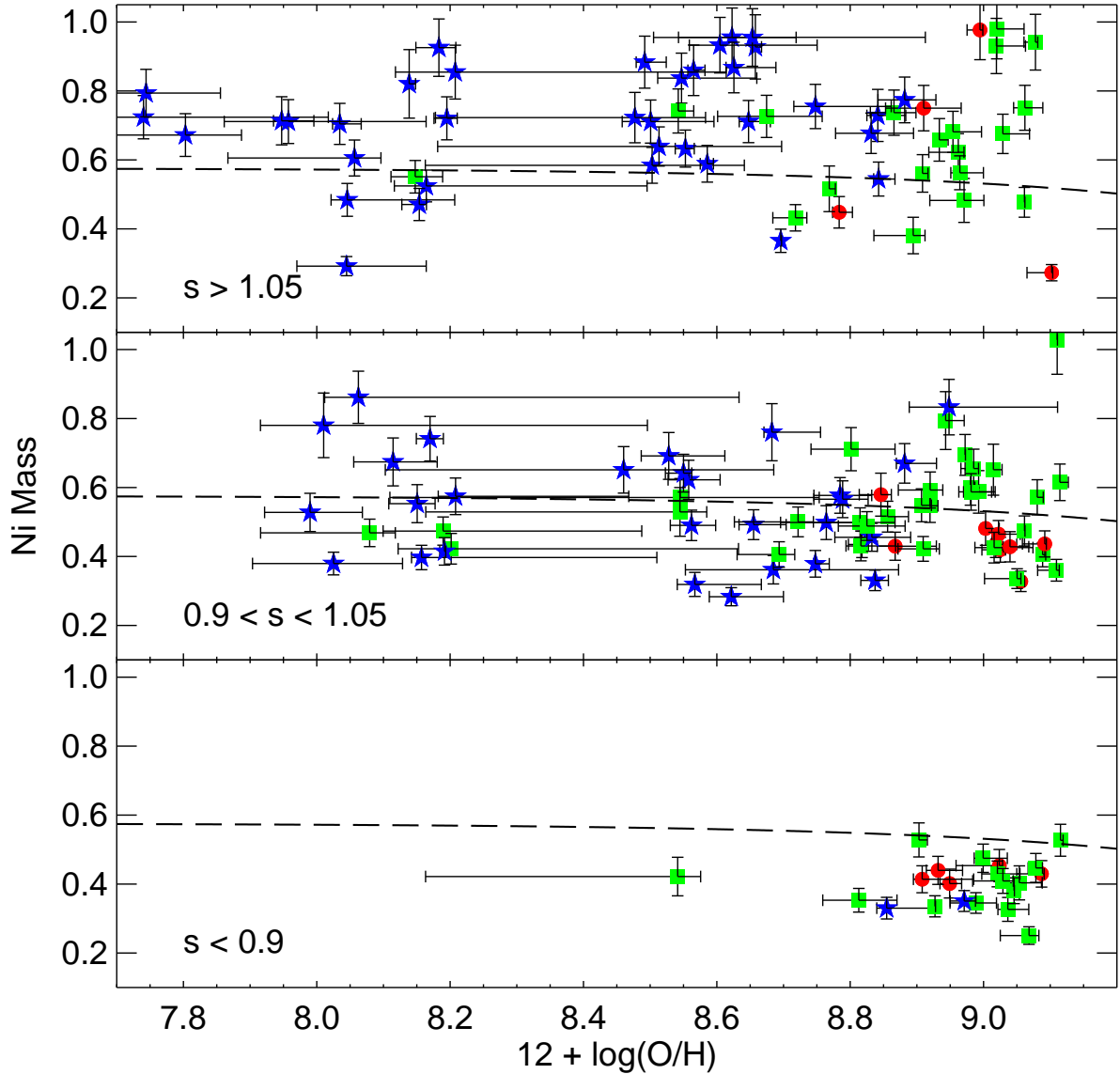} 
\caption{Ni mass versus implied O/H as in Figure \ref{metal}, split by
  stretch.  Within a stretch bin there is no obvious inferred host galaxy metallicity
  dependency (line shows TBT03 prediction).  The bottom panel shows that low stretch SNe Ia
  are always produced in higher-metallicity galaxies, and the scatter
  in \nickel[56] mass is lower for low-stretch SNe Ia.
\label{ni3p}}
\end{figure}

Figure \ref{ni3p} shows \nickel[56] mass vs. 12+log(O/H) in three
stretch bins.  The most striking finding is that low stretch ($s <
0.9$) SNe Ia almost exclusively occur in galaxies with high implied metallicity, roughly
above-solar.  At the highest stretches ($s > 1.05$), the
inferred \nickel[56] mass is relatively insensitive to inferred host galaxy
metallicity.  Also, the scatter in \nickel[56] mass is
lower for low stretch SNe Ia.  This is likely due to several factors.
For the low-stretch SNe, on the high side the \nickel[56] mass is
limited by the stretch cut, and on the low side it is limited by
selection effects -- it is difficult to find and spectroscopically
confirm the lowest-stretch SNe.  But there is also some evidence that
SNe Ia in passive galaxies, which predominantly host low stretch SNe
Ia, make better standard candles \citep{2003MNRAS.340.1057S}.  Some of
this lower dispersion is likely due to lower dust extinction in these
hosts, but it is also possible that moderately low-stretch SNe Ia are
intrinsically more uniform as a population than high stretch SNe Ia
\citep{howell2004}.

\begin{figure} 
  \plotone{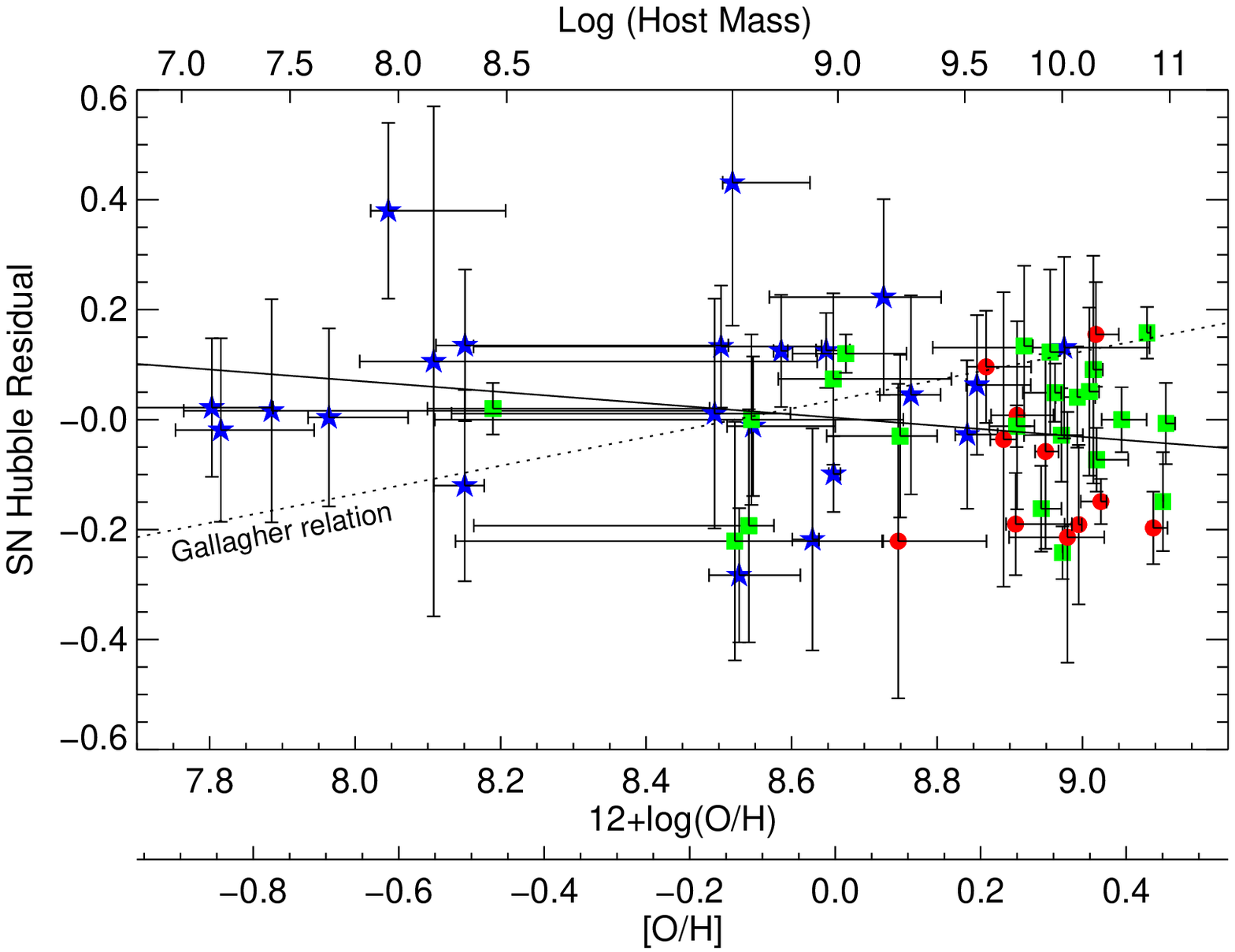} 
\caption{Hubble residual vs. inferred host galaxy metallicity for SNe Ia
  from the SNLS 1st year sample \citep{2006A&A...447...31A}.  The
  solid line is a fit to the data, though it is consistent with zero
  slope at the $\sim 1\sigma$ level.  The dotted line shows the
  G08 relation, here ruled out at $>99.9\%$
  confidence.  The metallicity dispersion of 0.1 dex in the
  \citet{2004ApJ...613..898T} relation was added in quadrature when
  calculating the fit, but is not plotted for clarity.  Residual
  errors do not include intrinsic SN Ia scatter about the Hubble line.
  No redshift or color cut was applied to the data; applying either
  or both gives a slope consistent with zero and still rules out the
  G08 relation at $>99.9\%$ confidence.
  Symbol shapes and colors are the same as Fig.~\ref{massstr}.  The
  upper axis shows the log mass for a galaxy with the metallicity
  found on the lower axis at $z=0$.  The convention used is the same
  as that in G08 --- overluminous SNe have a negative Hubble
  residual.  Note that we are using an indicator of [O/H] as the independent
  variable, whereas Gallagher use an indicator of [Fe/H].  Any
  systematic trend with Hubble residual should show a slope with
  respect to either variable.
\label{residmetal}}
\end{figure}

\subsection{Hubble residuals}\label{hubbleresidsec}
G08 found a trend between SN Ia host galaxy
metallicity and residuals from the Hubble diagram, indicating the
lightcurve fitter used, MLCS2k2, cannot correct for the full range of
SN Ia properties.  Since the demographics of supernovae are known to
evolve with redshift \citep{2007ApJ...667L..37H}, any problem fully
correcting SN Ia magnitudes could have dire consequences for SN Ia
cosmology.  \citeauthor{2008arXiv0805.4360G} cite a 9\% systematic
error in the measurement of the Dark Energy equation of state, $w$,
larger than the current best statistical errors on $w$
\citep{2008arXiv0804.4142K,2006A&A...447...31A}.  

However, using our independent metallicity indicator and the SiFTO
lightcurve fitter we find no systematic trend in Hubble residual
vs. inferred host galaxy metallicity.  The results are shown in Figure~\ref{residmetal},
where the SNe Ia include those from \citet{2006A&A...447...31A}
sample, but we apply no redshift or color cut.  The solid line shows
the best fit to the data, based on a Markov Chain Monte Carlo
analysis \citep[LINMIX in IDL;][]{2007ApJ...665.1489K}, which takes into account errors on the
x-axis, and is allowed to add extra scatter to the data (here 0.11
mag) to achieve the best fit.  The slope of the best-fit line, $-0.10\pm
0.07$ is consistent with 0 at the $\sim 1\sigma$ level, and is
inconsistent with the G08 slope of 0.26
at $5\sigma$.  In 10,000 Monte Carlo simulations, none produced a
slope as steep at that found by G08.  

Possible explanations for the difference between our results and those
of G08 include that SiFTO can correct for
differences in SN Ia properties that the current version of MLCS2k2 cannot, that one or the
other metallicity indicator produces systematically flawed results, or
that there are selection effects such that we are not comparing
similar samples.  We examine each hypothesis in turn.

{\bf SiFTO vs. MLCS: }  Recently \citet{2007ApJ...659..122J} found
evidence for a ``Hubble Bubble'' ---local SNe Ia fit with MLCS2k2 give
a different Hubble constant than do SNe Ia well into the Hubble flow.
\citet{2007ApJ...666..694W} considered this to be a 6\% systematic
error on $w$.  However, \citet{2007ApJ...664L..13C} did not see the
same effect with either the SiFTO or SALT2 \citep{2007A&A...466...11G}
lightcurve fitters, and traced the discrepancy to problems in the
way MLCS2k2 handled dust extinction.  MLCS2k2 assumed color excess
beyond any assumed intrinsic SN Ia color relation is due to $R_V=3.1$
(average Milky Way dust), though this appears not to be appropriate
(SALT2 and SiFTO do an empirical fit to determine color-luminosity
relationships).  G08 do not provide enough information to determine
whether a spurious reddening correction is driving their result.  

{\bf Flawed metallicity indicators: }  Since we cannot make a
definitive statement about whether the $y$-axis of
Figure~\ref{residmetal} (or the equivalent figure in G08) is flawed,
the next culprit to examine is the $x$-axis.  No metallicity indicator is
perfect --- the various caveats throughout the text attest to the
limitations of our adopted method.  Estimating a metallicity through a
mass-metallicity relationship and determining the mass from photometry
is certainly indirect.  Still, it should be correct on average,
especially since we are probing several orders of magnitude in mass,
ranging from $<10^8$ to $>10^{11}$, and more than 1 dex in
metallicity.  

A stronger argument is that there seems to be no way to rearrange the
points on the $x$-axis of Figure~\ref{residmetal} to produce the G08
relation.  On the other hand, it is possible to imagine that the $x$-axis of
the upper left panel of G08 Fig. 9 could be systematically flawed.
The four points with the highest metallicity drive the relation.
These four SNe all have E/S0 hosts with ages less than 3 Gyr, among
the lowest in the G08 sample.  Thus, the hosts of these SNe Ia are the most likely
to have H$\beta$ emission contaminating the H$\beta$ absorption line
index.  Nominally this is used to determine age, but because of the
age-metallicity degeneracy, it also affects the metallicity
determination.  As can be seen from G08 Figure 2, when the emission
correction is applied 1/3 to 1/2 of their sample either falls off or
is moved onto their theoretical age-metallicity grid.  Even after
correction, eight of their SNe Ia are more than $1\sigma$ too high off
the grid, and their metallicities are extrapolations.  

{\bf Selection effects: } Since G08 target nearby SNe Ia in E/S0
hosts, they study many SNe Ia too dim to show up in a
magnitude-limited sample like the SNLS.  But the four SNe with
high metallicities driving the G08 Hubble residual trend are all of
normal magnitude $-19 <M_V<-19.5$, not the type absent from the SNLS
sample.  The metallicity range is also similar for the two studies,
spanning a range $\pm \sim 0.5$ dex relative to solar.  But if the differences between
the two studies were due to some unknown selection effect, an
important point is that it will not produce a systematic effect on
cosmology, because the SNLS sample reflects cosmology as it is
actually practiced, and we have shown that it produces unbiased Hubble
residuals.  

\section{Conclusions}
We find that host galaxies with a higher inferred metallicity
produce SNe Ia with less \nickel[56] on average.  The effect proposed by
TBT03, increased neutronization in higher metallicity environments,
can qualitatively explain this trend, though there remains additional scatter in SN
Ia luminosity that cannot be explained by the theory. 
TBT03 predict that SNe Ia in high metallicity environments should be
25\% dimmer than those in low metallicity galaxies, though most of
the dynamic range in this prediction occurs at metallicities several times
solar.  We have shown that SNLS supernovae occur in galaxies where the
implied average metallicity is a few tenths of a dex above solar or below, and
in this range the theory predicts only a 0.06 \Msun\ change in
\nickel[56] mass.  The {\em average} \nickel[56] yield does appear to
decrease in metal-rich galaxies in accordance with the theory, but the
large scatter  (0.2 -- 1.0 \Msun , or $\sim0.6$ \Msun\ with
color correction), indicates that the theorized metallicity
effect has at most a 7\%--10\% effect on \nickel[56] yield.  

The luminosity
of SNe Ia appears to be dominated by other effects, possibly those
caused by the age of the progenitor \citep{2001ApJ...554L.193H,
  2005ApJ...634..210G, 2005A&A...433..807M, 2006ApJ...648..868S}.
We find a correlation between host galaxy luminosity-weighted age and
SN \nickel[56] yield.  This indicates that more massive stars
(explicitly, massive secondaries, since they set the timescale for
explosion in binaries), produce more luminous explosions, a result not
well understood if most SNe Ia explode near the Chandrasekhar mass. 
We have also shown that allowing for changes in O/Fe with the
metallicity [Fe/H] does not have a major effect on the theoretical
prediction of TBT03 for metallicities within the observed range of our
sample, so long as one is using the O/H as the
independent variable.

Unlike G08, we find no trends between residuals from the Hubble
diagram and host metallicity.  We conclude that metallicity is not a
significant systematic error for the measurement of $w$.  The G08
finding may be due to problems with their lightcurve fitter, MLCS2k2,
or with their method of host metallicity determination which relies on
an uncertain correction for H$\beta$ emission, and the extrapolation
of theoretical models to cover the host galaxies of SNe Ia which fall off the grid. 

These findings are broadly consistent with the range of local SN Ia
host galaxy metallicities found by \citet{2005ApJ...634..210G}, though
we differ in how we compare the model to the data, and in the
interpretation of the results.  Despite using completely different
methodology and data, both studies find that the majority of SN hosts
have a metallicity between 8.4 $<$ log (O/H) + 12 $<$ 9.1.
Additionally we find several lower metallicity, low mass hosts.  These
are underrepresented in low redshift searches which target large,
higher-mass galaxies \citep[e.g.][]{2001ApJ...546..734L}.  But
rather than convert the observations to \nickel[56] mass, 
\citet{2005ApJ...634..210G} converted the TBT03 model
prediction into an expected difference in \drp\ 
\citep{1993ApJ...413L.105P}.  They found a fairly steep evolution in
the \drp\ model prediction over the observed range in host O/H, in
contrast to the flat prediction over the same range found here
(Fig. \ref{metal}) and in TBT03.  This is because they used the
delayed detonation models of \citet{2002ApJ...568..791H} and the
empirical relations of \citet{2004ApJ...613.1120G} to convert the
model to the observed parameter \drp .  The
\citet{2002ApJ...568..791H} models had a varying detonation transition
density, so \citet{2005ApJ...634..210G} found a steeper
relation than one would derive if only O/H were varied while holding
the other properties of the explosion fixed.  As a result, they concluded
that the TBT03 prediction is a poor fit to the data, whereas we find
that it could account for some of the variation in SN Ia properties, but
it is not the dominant effect.

Our findings are also consistent with \citet{2008ApJ...673.1009P}, who
roughly estimated that the TBT03 theory could only explain a small
fraction of SN Ia luminosity based on average SNLS SN Ia stretches
reported in \citet{2007ApJ...667L..37H} and average SN host
metallicity estimated from the galaxy types in
\citet{2006ApJ...648..868S}.

One caveat regarding our findings is that we are using a proxy for the
global average galaxy metallicity -- we have no way of knowing
the actual progenitor's metallicity.  One could imagine an effect that
would skew the results, such as an enhanced SN rate in metal-poor or
metal-rich environments, or similar diminution of the rate in either
environment, such that the average SN metallicity would not be
representative of the average metallicity in the galaxy.  While
there has been speculation from time to time that such effects could
in principle exist \citep{1998ApJ...503L.155K,
  2000A&A...362.1046L}, there is no observational evidence that they
do.  

The absolute calibration of the mass-metallicity relationship is also
uncertain \citep{2008arXiv0801.1849K}, but changing it would have the
main effect of moving all points in Fig.~\ref{metal} to slightly lower
or higher metallicities.  Our conclusions would remain unchanged --
that the metallicity effect can only account for a fraction of the
scatter in SN Ia luminosities.


\citet{2008ApJ...673.1009P} find that weak reactions during low-level
burning in the white dwarf in the $\sim 1000$ years leading up to explosion
(i.e. simmering) may increase the neutron excess.  This effect may set
a ``floor'' to the level of neutronization that dominates over the
TBT03 metallicity effect at low metallicities $Z/Z_{\odot}\lesssim
2/3$.  \citet{2008arXiv0801.1643C} also find that simmering can
increase neutronization.  Since neutronization effects during
simmering depend on the degree of simmering, and at least to first
order do not appear to depend on metallicity, it is possible that
simmering could wash out the TBT03 metallicity effect.  Significant
uncertainties, including the role of the convective Urca process,
remain, and further study is required.  At the very least it seems
that simmering cannot enhance the TBT03 metallicity effect enough to
account for the significant scatter in SN Ia luminosities
\citep{2008ApJ...673.1009P}.

One additional caveat is that it is possible that a change in
metallicity could alter the explosion mechanism itself, not just
whether the process produces more \nickel[56] or \nickel[58].
\citet{2007ApJ...655L..93C} found that laminar flame speed also
depends on the metallicity, but this is probably a higher-order
effect.  At any rate, this process is so poorly understood that it is
not yet possible to test this effect with observations.

\acknowledgements

The authors thank Lars Bildsten, Kevin Bundy, and
Renbin Yan for helpful discussions.  We also thank the Kavli
Institute for Theoretical Physics, and the Aspen Center for Physics, 
where the initial development of
this work was done.  This research was supported in part by the
National Science Foundation under Grant No. PHY05-51164.

EFB is supported by grant AST-0507456 from the National Science
Foundation.  MS acknowledges support from the Royal Society.  JCW is supported in part by NSF AST-0707769

The SNLS collaboration gratefully acknowledges the assistance of Pierre
Martin and the CFHT Queued Service Observations team. 
Canadian collaboration members acknowledge support from NSERC and
CIAR; French collaboration members from CNRS/IN2P3, CNRS/INSU and CEA.

SNLS relies on observations with MegaCam, a joint project of
CFHT and CEA/DAPNIA, at the Canada-France-Hawaii Telescope (CFHT)
which is operated by the National Research Council (NRC) of Canada, the
Institut National des Science de l'Univers of the Centre National de la
Recherche Scientifique (CNRS) of France, and the University of Hawaii. This
work is based in part on data products produced at the Canadian
Astronomy Data Centre as part of the Canada-France-Hawaii Telescope Legacy
Survey, a collaborative project of the National Research Council of
Canada and the French Centre national de la recherche scientifique.

\bibliographystyle{apj}

\end{document}